\documentclass[
]{ceurart}

\usepackage{listings}
\usepackage{placeins}
\usepackage{caption}
\usepackage{subcaption}
\usepackage{hyperref}
\usepackage{float}
\begin{document}

\copyrightyear{2026}
\copyrightclause{Copyright for this paper by its authors.
  Use permitted under Creative Commons License Attribution 4.0
  International (CC BY 4.0).}

\conference{Late Interaction Workshop (LIR) @ ECIR 2026, April 02, 2026. Colocated with ECIR 2026.}

\title{Working Notes on Late Interaction Dynamics: Analyzing Targeted Behaviors of Late Interaction Models}

\author[1]{Antoine Edy}[%
email=antoine.edy@illuin.tech,
]
\cormark[1]
\fnmark[1]
\address[1]{Illuin Technology}
\address[2]{CentraleSupélec, Paris-Saclay}

\author[1]{Max Conti}[%
email=max.conti@illuin.tech,
]
\fnmark[1]

\author[1, 2]{Quentin Macé}[%
email=quentin.mace@illuin.tech,
]
\fnmark[1]

\cortext[1]{Corresponding author.}
\fntext[1]{These authors contributed equally.}

\newcommand{\aedy}[1]{{\textcolor{teal}{\bf [{\sc AE:} #1]}}}
\newcommand{\mconti}[1]{{\textcolor{olive}{\bf [{\sc MC:} #1]}}}
\newcommand{\qmace}[1]{{\textcolor{orange}{\bf [{\sc QM:} #1]}}}
\newcommand{\vx}[1]{{\textcolor{blue}{\bf [{\sc VX:} #1]}}}

\begin{abstract}
While Late Interaction models exhibit strong retrieval performance, many of their underlying dynamics remain understudied, potentially hiding performance bottlenecks.
In this work, we focus on two topics in Late Interaction retrieval: a length bias that arises when using multi-vector scoring, and the similarity distribution beyond the best scores pooled by the MaxSim operator. We analyze these behaviors for state-of-the-art models on the NanoBEIR benchmark. Results show that while the theoretical length bias of causal Late Interaction models holds in practice, bi-directional models can also suffer from it in extreme cases. We also note that no significant similarity trend lies beyond the top-1 document token, validating that the MaxSim operator efficiently exploits the token-level similarity scores.
\end{abstract}


\maketitle

\section{Introduction} 

Neural late-interaction retrieval models, such as ColBERT \cite{khattab2020colbertefficienteffectivepassage}, use a token-level interaction while computing similarity between text passages. While this approach allows for finer semantic matching between queries and documents, some of its underlying dynamics are yet to be thoroughly studied.

In these notes, we analyze two key behaviors that provide elements for a better understanding of Late Interaction performance:

\noindent\textbf{(a) Length bias:} Causal encoders, when used with multi-vector MaxSim scoring, exhibit a monotonic bias that favors longer chunks, regardless of their true relevance. 

\noindent\textbf{(b) Similarity distribution:} Given a query token, the MaxSim operator is insensitive to similarity scores of document tokens beyond the highest, collapsing information to a single maximum value.

We perform small-scale experiments on the NanoBEIR \cite{nanobeir} benchmark to bring more insights into how current state-of-the-art models behave along these axes. 

\section{Length Bias In Multi-Vector Retrieval}

In this section, we explore the length bias that arises in late-interaction retrieval frameworks. We highlight two key observations: multi-vector causal models appear to suffer from a strict, monotonic length bias, and while bi-directional architectures theoretically avoid this flaw, empirical insights suggest they remain sensitive to length disparities at extreme margins. Details on the experimental setup are available in \autoref{app:exp_setup}.

\subsection{Theoretical Motivation}

Let a chunk $c$ be a sequence of tokens represented by contextualized embeddings. In late-interaction retrieval, the MaxSim score \cite{khattab2020colbertefficienteffectivepassage} between a query $q$ and a chunk $c$ is defined as:

$$S_{q,c} = \sum_{i \in [|E_q|]} \max_{j \in [|E_c|]} E_{q_i} \cdot E_{c_j}^T$$ \label{eq:max_sim}

where $E_q$ and $E_c$ are the respective sets of query and chunk embeddings. When utilizing a causal encoder with a multi-vector representation, appending tokens to a chunk yields a strict superset of embeddings. As a result, the maximum inner product for each query token can only increase or remain constant. This dynamic introduces a theoretical monotonic length bias that artificially favors longer chunks regardless of their true relevance.

Bi-directional and single-vector models theoretically avoid this strict bias. In bi-directional models, appending new tokens alters the attention context for all preceding tokens, allowing scores to naturally decrease if the semantic focus is diluted. Similarly, single-vector models aggregate tokens into a fixed-length representation that does not inherently benefit from added tokens. However, to understand their practical robustness to length differences, we complement this theoretical intuition with an empirical analysis.

\subsection{Multi-Vector Architectures Induce A Length Bias} \label{sec:length-bias-causal}

\begin{figure}[!h]
  \centering
  \includegraphics[width=\textwidth]{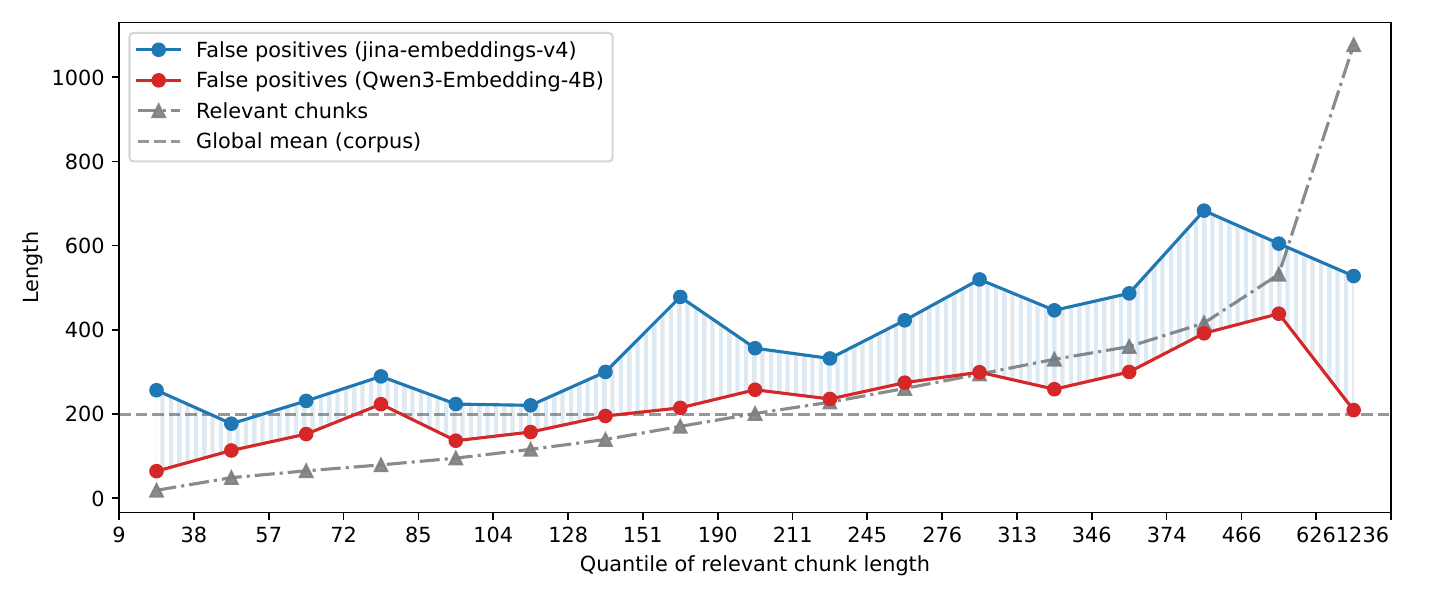}
  \caption{Mean length comparison between the retrieved false positive chunks, the relevant ground-truth documents, and the global corpus average. Queries are grouped into quantiles on the x-axis based entirely on the average token length of their corresponding relevant chunks.}
  \label{fig:enter-label}
\end{figure}

Figure \ref{fig:enter-label} isolates the impact of the pooling mechanism in causal architectures by comparing a multi-vector model (\texttt{jina-embeddings-v4} \cite{günther2025jinaembeddingsv4universalembeddingsmultimodal}) with a single-vector dense model (\texttt{Qwen3-Embedding-4B} \cite{qwen3embedding}). Queries on the horizontal axis are partitioned into quantiles based on the token length of their true relevant documents. For each quantile, we compare the mean length of retrieved false positives against the true relevant documents, as well as the global corpus mean (199 tokens).

While false positives are statistically expected to be slightly longer on average due to the inherently wider semantic scope of longer texts, this length gap should ideally remain marginal. However, the multi-vector causal model disproportionately retrieves false positives that are significantly longer than the relevant documents. Conversely, the single-vector causal model tracks the relevant document length much more closely. This confirms that within causal architectures, the multi-vector setup is the primary driver of length bias.

\subsection{Bi-Directional Models Mitigate But Do Not Eliminate Bias} \label{sec:length-bias-bidir}

\begin{figure}[h!]
  \centering
  \begin{subfigure}[t]{0.48\textwidth}
    \centering
    \includegraphics[width=\textwidth]{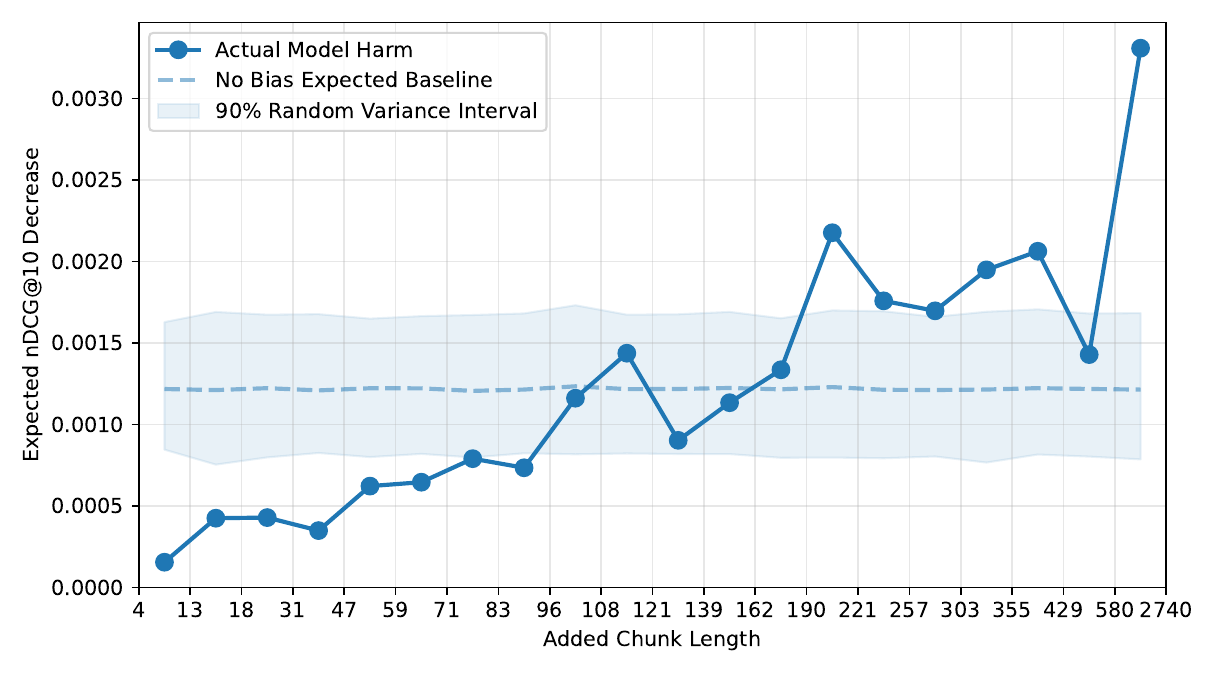}
    \caption{\texttt{jina-embeddings-v4}}
    \label{fig:harm_jina}
  \end{subfigure}\hfill
  \begin{subfigure}[t]{0.48\textwidth}
    \centering
    \includegraphics[width=\textwidth]{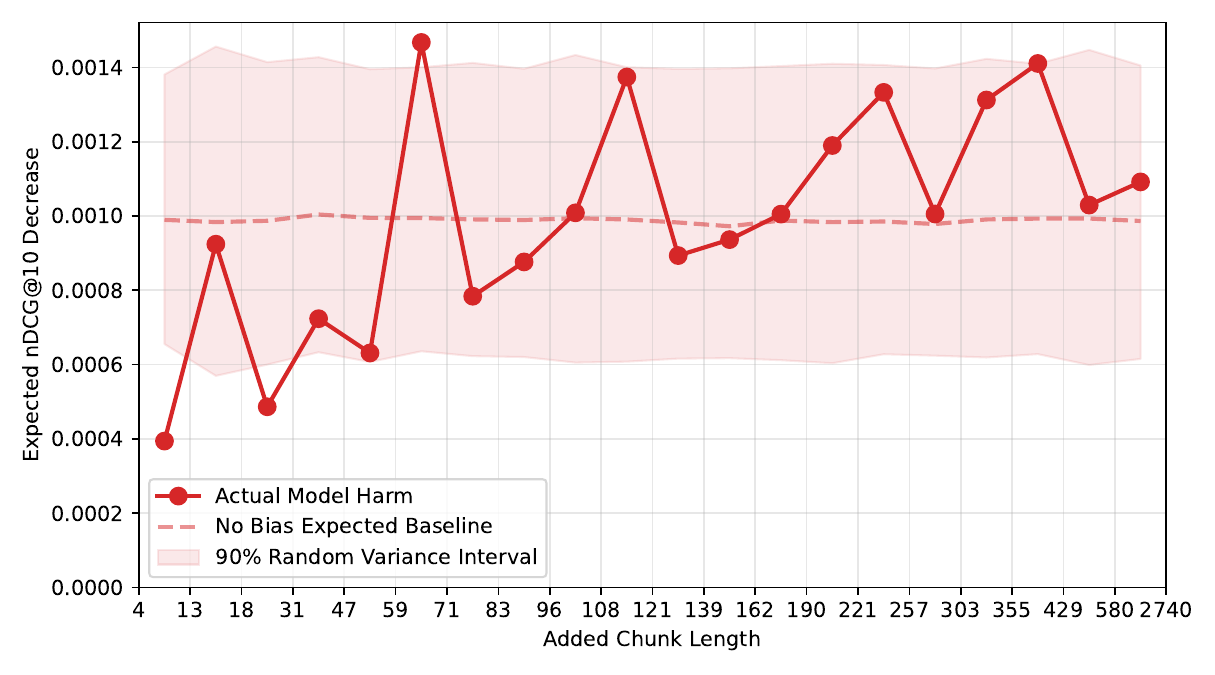}
    \caption{\texttt{Qwen3-Embedding-4B}}
    \label{fig:harm_qwen}
  \end{subfigure}
  \begin{subfigure}[t]{0.48\textwidth}
    \centering
    \includegraphics[width=\textwidth]{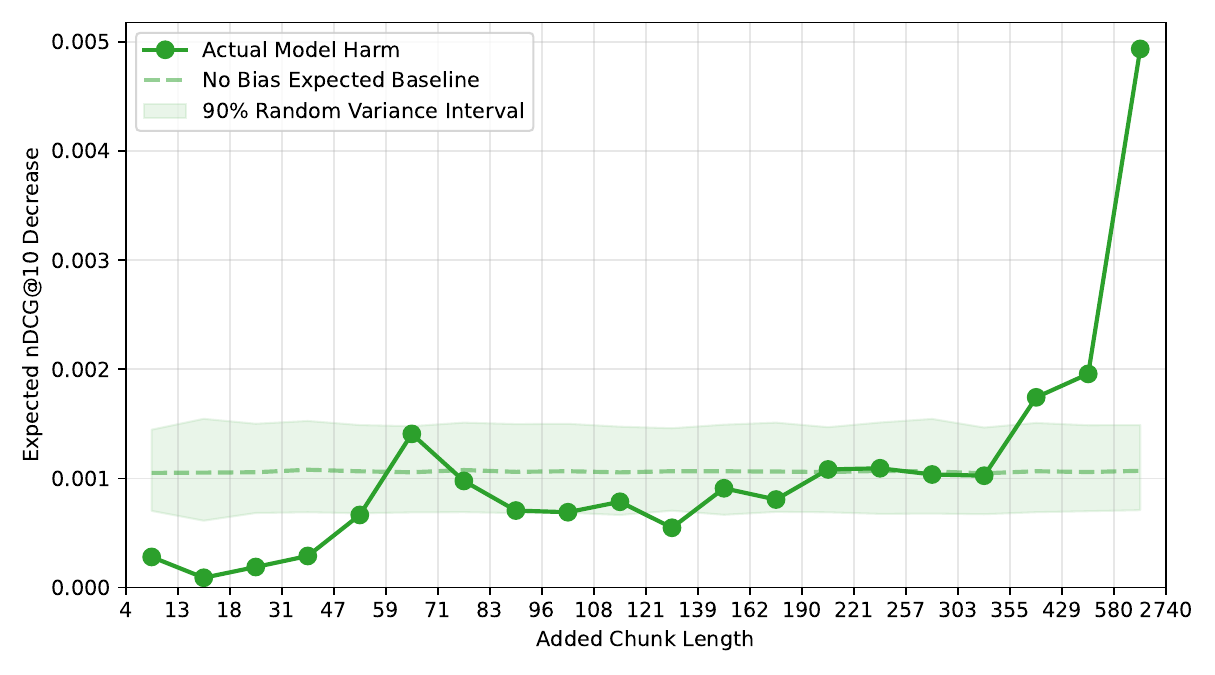}
    \caption{\texttt{GTE-ModernColBERT-v1}}
    \label{fig:harm_moderncolbert}
  \end{subfigure}\hfill
  \begin{subfigure}[t]{0.48\textwidth}
    \centering
    \includegraphics[width=\textwidth]{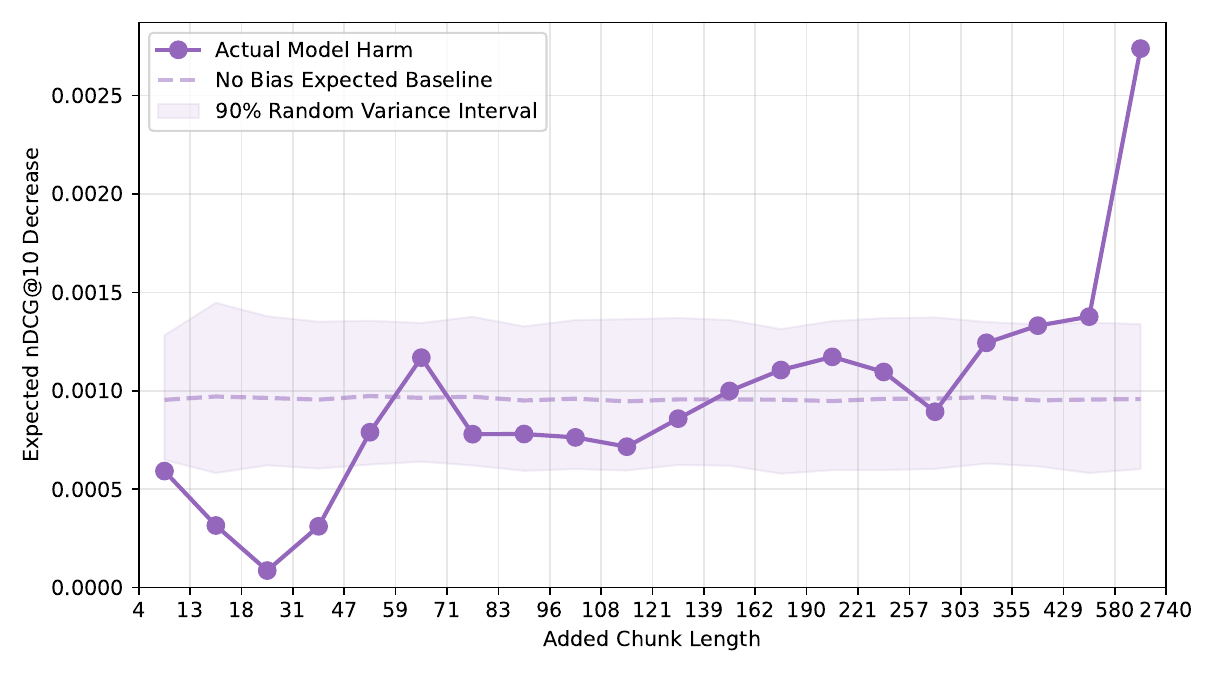}
    \caption{\texttt{ColBERT-Zero}}
    \label{fig:harm_colbert_zero}
  \end{subfigure}
  \caption{Expected decrease in retrieval performance (nDCG@10) when a chunk of a specific length is added to the corpus. Chunks are categorized into equal-sized quantile bins by token length on the x-axis. The solid line plots the average nDCG penalty incurred by the presence of a chunk from that bin, evaluated against a random baseline (dashed line) and its 90\% confidence interval (shaded area).}
  \label{fig:harm_effect_4_figures}
\end{figure}

Having identified the multi-vector setup as a primary driver of length bias in causal architectures, we now investigate whether bi-directional attention mechanisms can effectively neutralize this issue. We evaluate the expected unconditional decrease in retrieval performance (nDCG) when a chunk of a specific length is added to the corpus. To isolate the effect of length, we compute a random baseline and a 90\% confidence interval using a permutation test that assumes no correlation between chunk length and retrieval harm. Any deviation outside this interval demonstrates a statistically significant length bias, indicating that adding chunks of that size disproportionately harms ranking quality compared to random chance.

The results reveal distinct architectural behaviors. The causal multi-vector model (Figure \ref{fig:harm_jina}) exhibits a near-monotonic length bias; adding longer chunks consistently translates to a greater expected decrease in nDCG. In contrast, the single-vector dense model (Figure \ref{fig:harm_qwen}) displays no significant length bias, remaining safely within the random baseline and corroborating the findings of the previous section. Notably, while bi-directional multi-vector models successfully dampen the aggressive bias of causal models, they remain vulnerable at length extremes (Figures \ref{fig:harm_moderncolbert} and \ref{fig:harm_colbert_zero}). For these models, adding unusually short chunks is significantly less harmful than expected by random chance, whereas introducing exceptionally long chunks disproportionately degrades overall ranking quality. Thus, while bi-directional attention refines token representations, it remains unable to fully recover the systemic length bias introduced by the MaxSim operation across substantial token variations.

\FloatBarrier

\section{Similarity Distribution: What Happens Beyond The Top-1 Document Token} 

The MaxSim operator described in \autoref{eq:max_sim}, by construction, only considers the single most similar document token for each query token, ignoring the number and density of relevant tokens in a chunk. 
This could lead to hypersensitivity to the highest similarity score of document tokens. Intuitively, a single query token that has several strong matches in a document A is more similar to it than to a document B, where only one token has a high similarity, a nuance that is lost during the MaxSim operation.

To understand how document token similarity behaves beyond the top-1, we focus on queries where retrieval fails (i.e., when the positive document is outside of the top-10), to understand whether there are unseen similarity trends that could be exploited through alternative functions. We compare the sorted document token scores on these queries, aggregated over all query tokens, and across the failing queries on the NanoBEIR dataset, for both \texttt{ColBERT-Zero} and \texttt{jina-embeddings-v4}. We compare the similarity distribution for the positive sample that was not retrieved, and multiple negatives: the top-1 (best), the one ranked directly below the positive, and the worst.

\begin{figure}[htbp]
    \centering
    \begin{subfigure}[b]{0.49\textwidth}
        \centering
        \includegraphics[width=\linewidth]{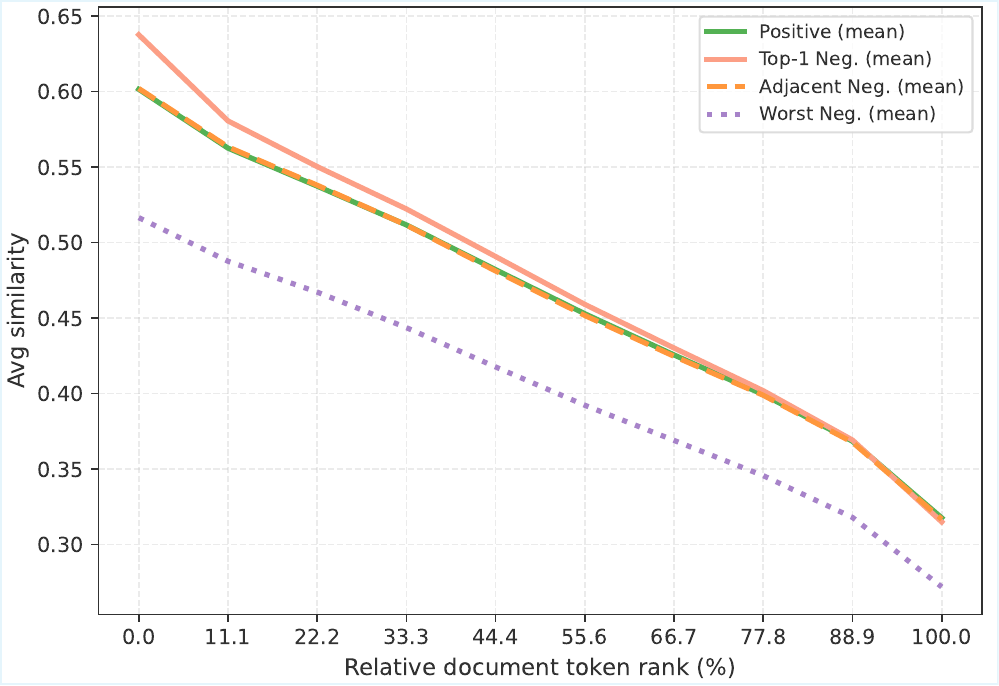}
        \caption{Similarities aggregated across datasets (201 queries).}
        \label{fig:colbert_agg}
    \end{subfigure}
    \hfill
    \begin{subfigure}[b]{0.49\textwidth}
        \centering
        \includegraphics[width=\linewidth]{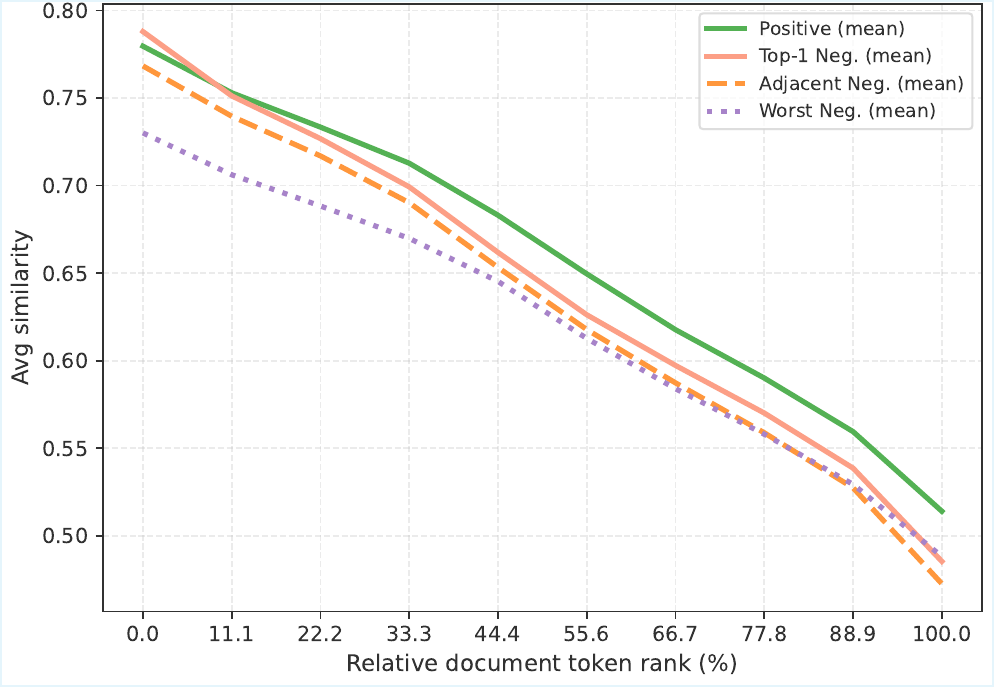}
        \caption{Similarities on NanoArguAna (15 queries).}
        \label{fig:colbert_nanoarguana}
    \end{subfigure}
    
    \caption{\texttt{ColBERT-Zero} document token similarities on failed queries. While some datasets exhibit interesting results (e.g., NanoArguAna, right), no clear tendency emerges for positive documents across NanoBEIR (left).
    }
    \label{fig:sim_distrib_comparison}
\end{figure}

As \autoref{fig:colbert_nanoarguana} shows, on NanoArguAna, the positive document has better document similarity than the top-1 negative beyond the first tokens (starting around 10\%). In such cases, leveraging score distribution beyond the best document token could help identify positive chunks. However, this result does not hold on average across all datasets of NanoBEIR, suggesting that such techniques would not generalize well enough. We also analyze the behavior on successful retrieval samples, with similar conclusions: positive retrieved samples do not have a significantly different similarity distribution in document tokens than negatives.
\texttt{jina-embeddings-v4} exhibits the same trends (\autoref{app:jina_similarity}), that further hold across all individual datasets.

\section{Conclusion And Future Work}

This work provides insights into Late Interaction model dynamics, highlighting a strict length bias in multi-vector causal architectures that bi-directional models mitigate. Similarly to previous work \cite{teiletche2025modernvbertsmallervisualdocument}, this suggests that causal models are a poor fit for Late Interaction, encouraging the development of stronger bi-directional models trained for this paradigm. Furthermore, we observed that on standard retrieval benchmarks, no significant similarity trends emerge beyond the top-1 document token, suggesting that current models do not provide information that could be leveraged beyond a MaxSim operator.

Interesting future analyses include testing for a length bias in a controlled setting, using synthetic data to precisely adjust text lengths and semantic relevance. Additionally, document token score distributions should be analyzed on a broader set of tasks, to see if task-dependent trends could emerge (e.g., in long-context retrieval). It remains an important analysis to do on newly released models, as training recipes can highly impact these behaviors. 

Future work could also explore translating these insights to mitigate bias and enhance performance through interventions at training time, during indexing, or by refining the retrieval similarity operator.

\bibliography{bib}

\FloatBarrier

\newpage

\section*{Appendix}

\appendix

\section{Experimental Setup}\label{app:exp_setup}

\subsection{Datasets}
To empirically validate length bias, we evaluated retrieval performance using NanoBEIR \cite{nanobeir}, a subset of the BEIR benchmark \cite{thakur2021beirheterogenousbenchmarkzeroshot} comprising 13 diverse datasets with 50 queries each. To ensure a wide distribution of chunk lengths, we pooled the 13 datasets into a single unified corpus prior to retrieval. After removing five outlier chunks exceeding 3,000 tokens (which lacked associated queries), the final merged corpus contained 56,718 chunks and 649 queries.

\begin{figure}[!ht]
    \centering
    \includegraphics[width=0.7\textwidth]{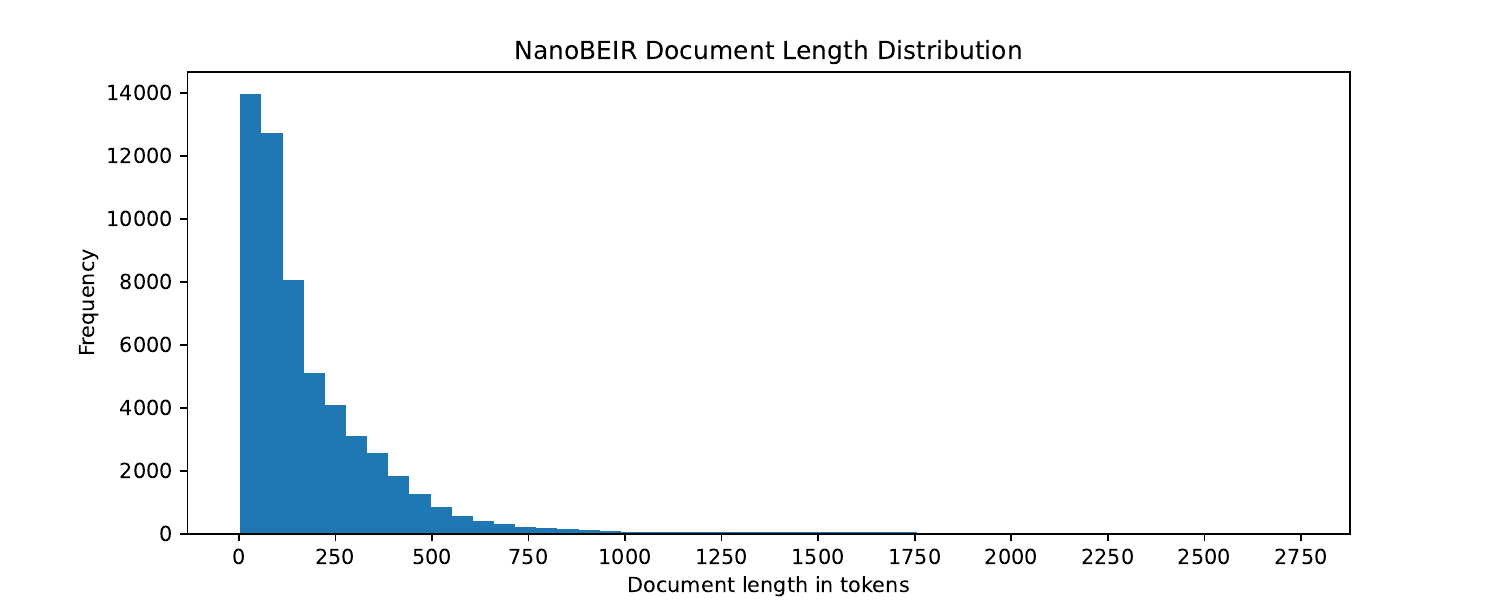}
    \caption{Chunk Length Distribution across the merged NanoBEIR corpus.}
    \label{fig:nanobeir-length-distrib}
\end{figure}

\subsection{Setup}
Chunk sizes were computed using the \texttt{jina-embeddings-v4} \cite{günther2025jinaembeddingsv4universalembeddingsmultimodal} tokenizer, a Byte-Pair Encoding tokenizer inherited directly from its base model, Qwen2.5-VL-3B-Instruct \cite{bai2025qwen25vltechnicalreport}. We evaluated four distinct model configurations representing combinations of encoder architectures (causal vs. bi-directional) and pooling strategies (single-vector vs. multi-vector):

\begin{table}[htpb]
    \centering
    \begin{tabular}{l c c c}
        \hline
        \textbf{Model} & \textbf{Pooling Strategy} & \textbf{Architecture} & \textbf{Parameters} \\
        \hline
        \texttt{jina-embeddings-v4} \cite{günther2025jinaembeddingsv4universalembeddingsmultimodal} & Multi-vector & Causal & 4B \\
        \texttt{Qwen3-Embedding-4B} \cite{qwen3embedding} & Single-vector & Causal & 4B \\
        \texttt{GTE-ModernColBERT-v1} \cite{GTE-ModernColBERT} & Multi-vector & Bi-directional & 0.15B \\
        \texttt{ColBERT-Zero} \cite{chaffin2026colbertzeropretrainpretraincolbert} & Multi-vector & Bi-directional & 0.15B \\
        \hline
    \end{tabular}
    \caption{Summary of the evaluated models, including their pooling strategies, architectures, and sizes.}
    \label{tab:model_configurations}
\end{table}

\section{Retrieval Errors by Chunk Length}

\begin{figure}[h!]
    \centering
    \begin{subfigure}[t]{0.48\textwidth}
        \centering
        \includegraphics[width=\textwidth]{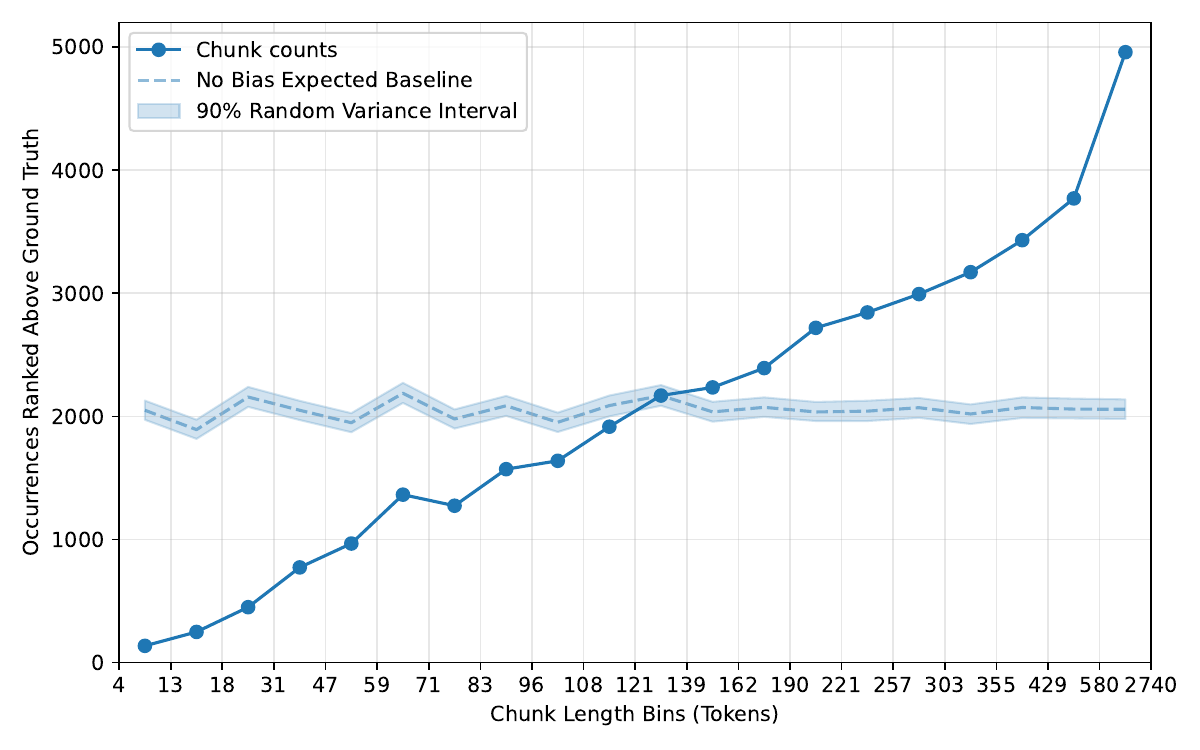}
        \caption{\texttt{jina-embeddings-v4}}
        \label{fig:length_jina}
    \end{subfigure}\hfill
    \begin{subfigure}[t]{0.48\textwidth}
        \centering
        \includegraphics[width=\textwidth]{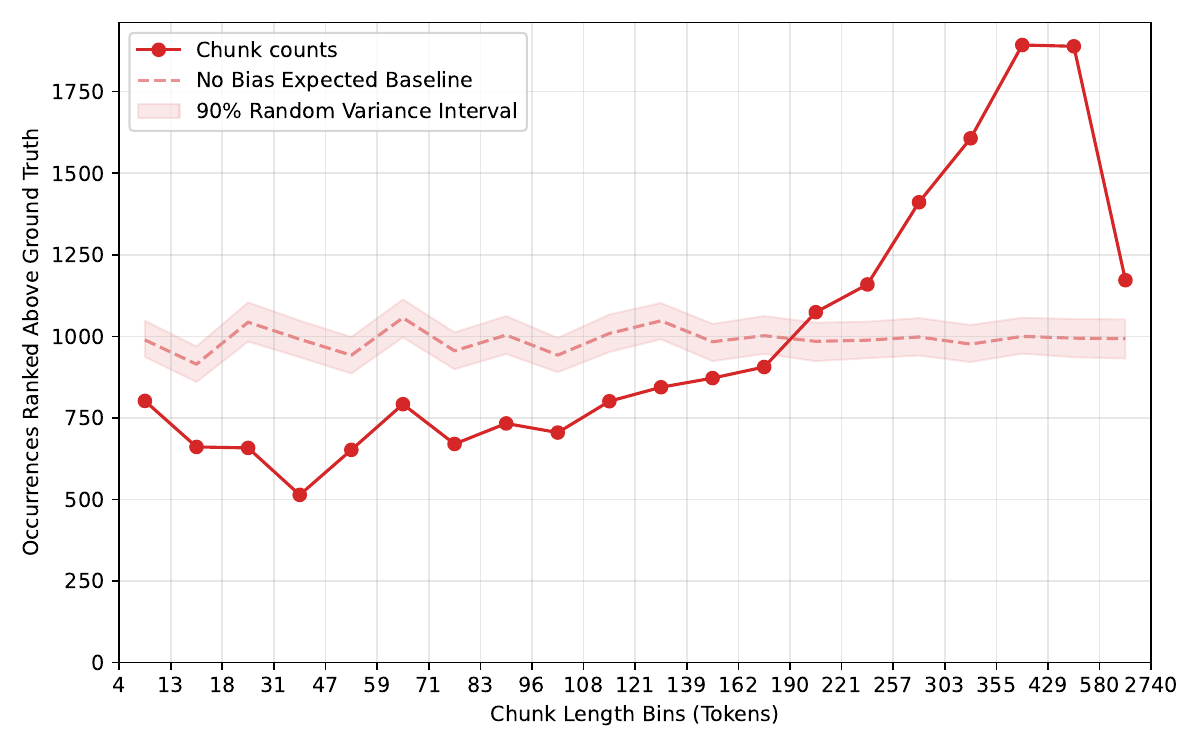}
        \caption{\texttt{Qwen3-Embedding-4B}}
        \label{fig:length_qwen}
    \end{subfigure}
    \begin{subfigure}[t]{0.48\textwidth}
        \centering
        \includegraphics[width=\textwidth]{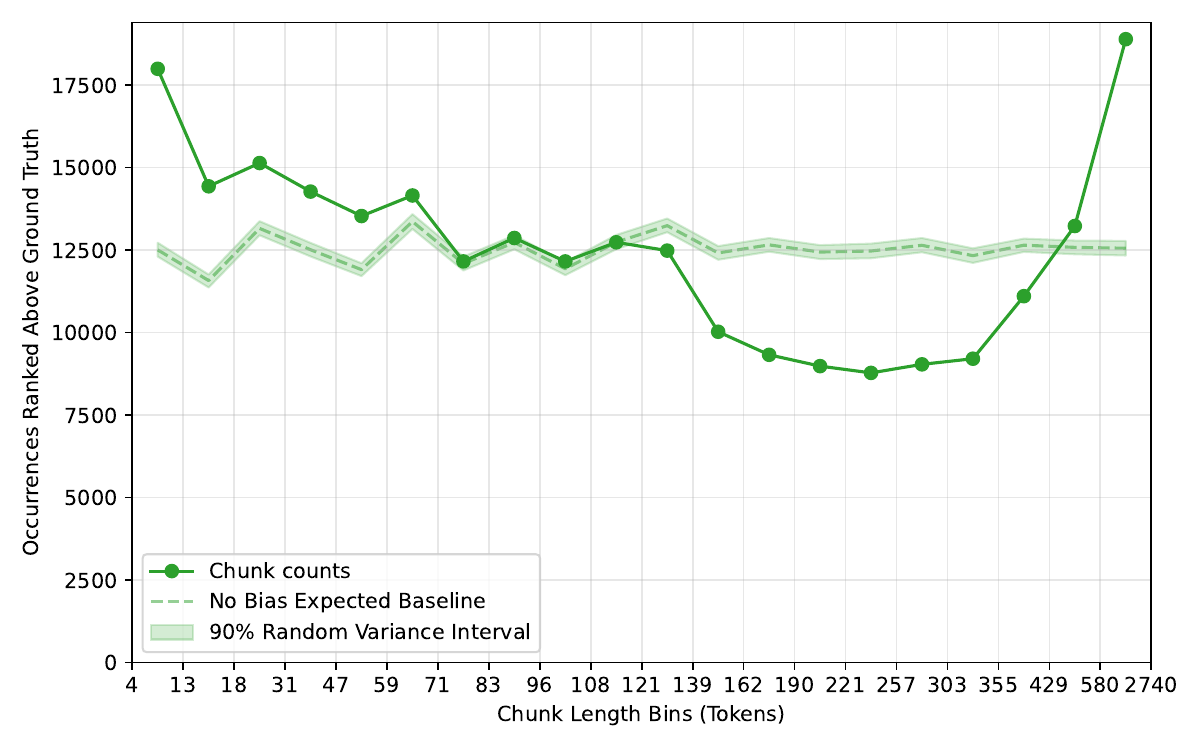}
        \caption{\texttt{GTE-ModernColBERT-v1}}
        \label{fig:length_moderncolbert}
    \end{subfigure}\hfill
    \begin{subfigure}[t]{0.48\textwidth}
        \centering
        \includegraphics[width=\textwidth]{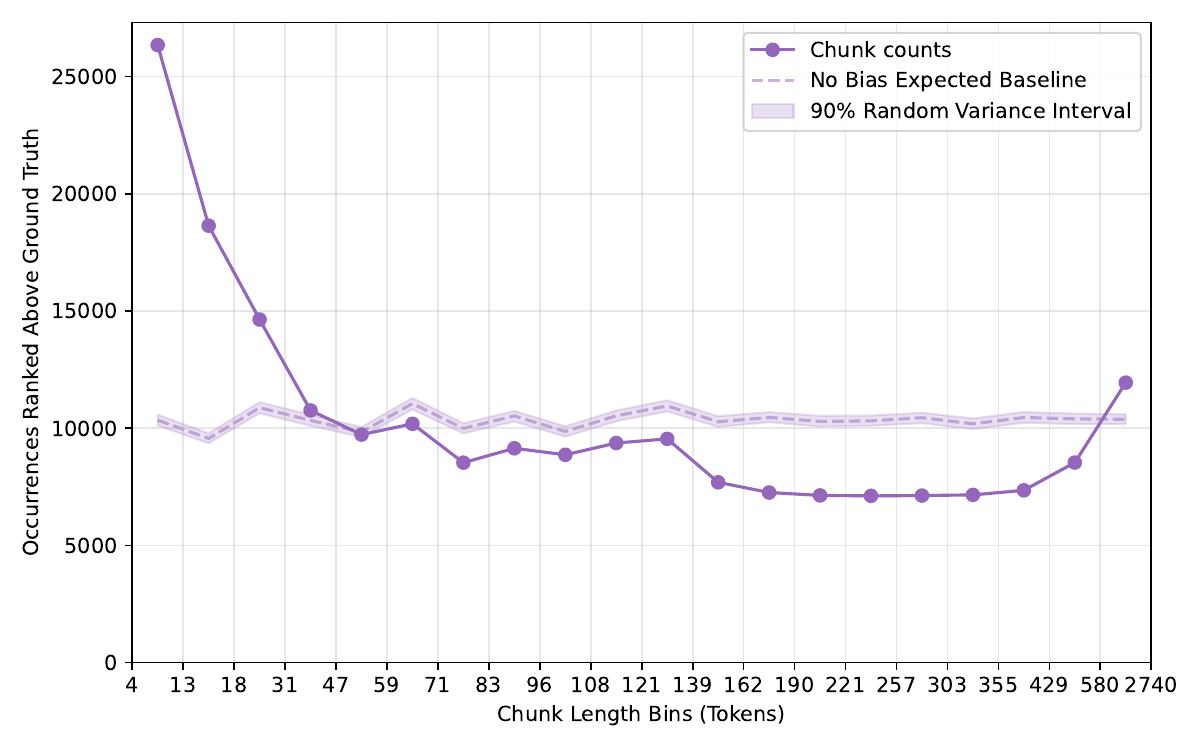}
        \caption{\texttt{ColBERT-Zero}}
        \label{fig:length_colbert_zero}
    \end{subfigure}
    \caption{Absolute occurrences of irrelevant chunks ranked above the highest-ranked true positive passage. Bin limits are defined to contain an equal number of chunks. The dashed line plots the no-bias expected baseline, bounded by a 90\% variance interval.}
    \label{fig:length_effect_4_figures}
\end{figure}

Figure \ref{fig:length_effect_4_figures} illustrates the raw volume of retrieval errors mapped to document chunk lengths.

While a general increase in false positives for longer chunks is expected as longer texts naturally contain more semantic coverage, the models exhibit distinct architectural vulnerabilities. The causal multi-vector model (\texttt{jina-embeddings-v4}) is the only configuration to display a strictly monotonic increase in errors starting from zero, corroborating the length bias of causal multi-vector models described in Section \ref{sec:length-bias-causal}. 

Conversely, bi-directional models demonstrate non-monotonic error distributions. Instead, distinct peaks emerge at the extreme ends of the length spectrum (for both very short and very long chunks), serving as an additional indicator of the marginal vulnerabilities discussed in Section \ref{sec:length-bias-bidir}. As a side note, both bi-directional models show a high absolute volume of errors despite yielding strong overall nDCG metrics, indicating that while they rank well on average, their failures are notably severe. This contrast in error severity can largely be explained by the substantial gap in parameter count compared to the two causal models (0.15B vs. 4B).

\FloatBarrier

\section{Similarity Distribution For jina-embeddings-v4}\label{app:jina_similarity}

\begin{figure}[htbp]
    \centering
    \begin{subfigure}[b]{0.49\textwidth}
        \centering
        \includegraphics[width=\linewidth]{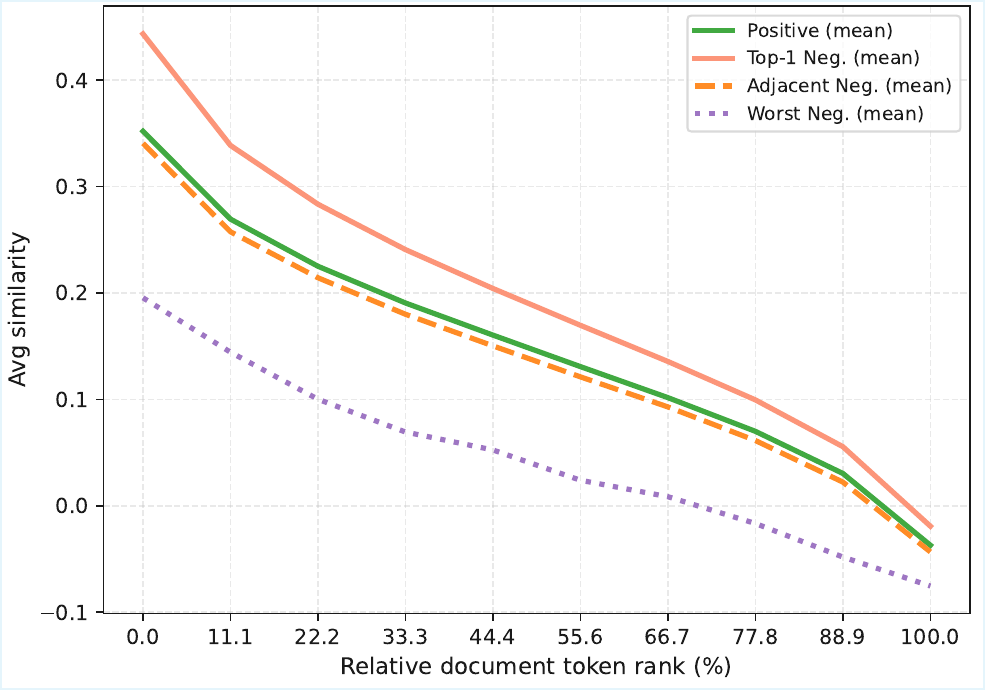}
        \caption{Similarities aggregated across datasets (229 queries).}
        \label{fig:jina_agg}
    \end{subfigure}
    \hfill
    \begin{subfigure}[b]{0.49\textwidth}
        \centering
        \includegraphics[width=\linewidth]{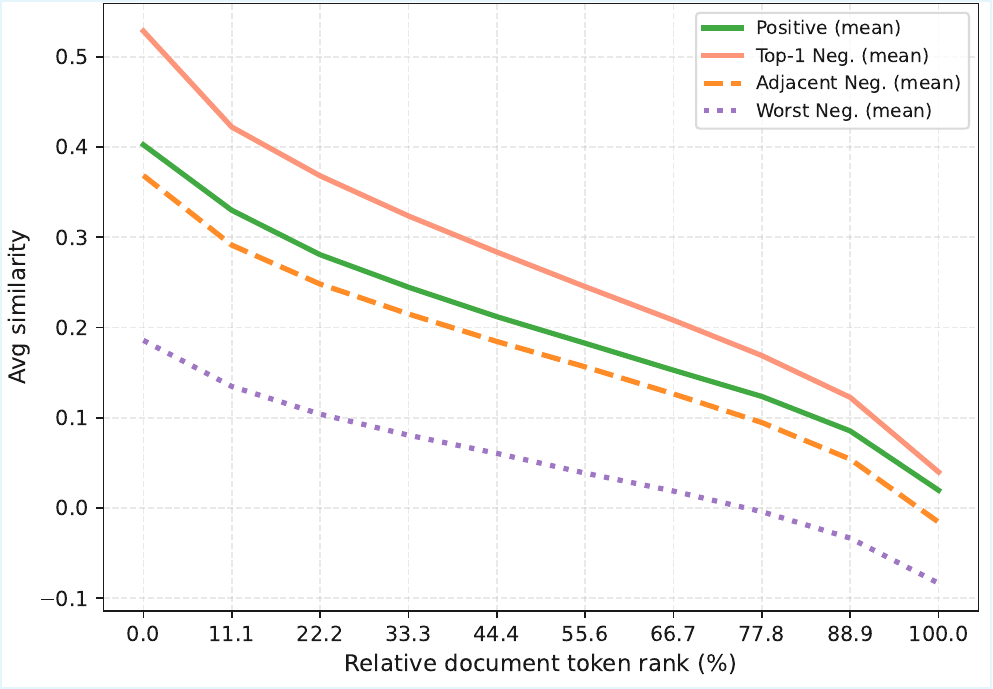}
        \caption{Similarities on NanoArguAna (5 queries).}
        \label{fig:jina_nanoarguana}
    \end{subfigure}
    
    \caption{\texttt{jina-embeddings-v4} document token similarities on failed queries. The positive document has a larger distance to the top-1 negative, and remains constantly worse across all document tokens.}
    \label{fig:jina_sim_distrib_comparison}
\end{figure}

\end{document}